\documentclass[preprint,superscriptaddress,showpacs,preprintnumbers,amsmath,amssymb]{revtex4}

\usepackage{graphicx}
\usepackage{dcolumn}
\usepackage{bm}
\usepackage{txfonts}

\def\lesssim{\ \raise.3ex\hbox{$<$}\kern-0.8em\lower.7ex\hbox{$\sim$}\ }
\def\gesim{\ \raise.3ex\hbox{$>$}\kern-0.8em\lower.7ex\hbox{$\sim$}\ }

\def\rnum#1{\expandafter{\romannumeral #1}} 
\def\Rnum#1{\uppercase\expandafter{\romannumeral #1}}

\begin{document}
\title{Strong-Coupling and Finite Temperature Effects on $p$-wave Contacts}

\author{Daisuke Inotani$^1$  and Yoji Ohashi$^1$}
 \affiliation{$^1$Department of Physics, Keio University, 3-14-1 Hiyoshi, Kohoku-ku, Yokohama 223-8522, Japan \\
}
\date{\today}

\begin{abstract}
We theoretically investigate strong-coupling and finite temperature effects on the $p$-wave contacts, as well as the asymptotic behavior of the momentum distribution in large momentum region in a one-component Fermi gas with a tunable $p$-wave interaction. Including $p$-wave pairing fluctuations within a strong-coupling theory, we calculate the $p$-wave contacts above the superfluid transition temperature $T_{\rm c}$ from the adiabatic energy relations. We show that  while the $p$-wave contacts related to the scattering volume monotonically increases with increasing the interaction strength, one related to the effective range non-monotonically depends on interaction strength and its sign changes in the intermediate-coupling regime. The non-monotonic interaction dependence of these quantities is shown to originate from the competition between the increase of the cutoff momentum and the decrease of the coupling constant of the $p$-wave interaction with increasing the effective range. We also analyze the asymptotic form of the momentum distribution in large momentum region. In contrast to the conventional $s$-wave case, we show that the asymptotic behavior cannot be completely described by only the $p$-wave contacts, and the extra terms, which is not related to the thermodynamic properties, appear. Furthermore, in high temperature region, we find that the extra terms dominate the sub-leading term of the large-momentum distribution. We also directly compare our results with the recent experimental measurement, by including the effects of a harmonic trap potential within the local density approximation. We show that our model explains the dependence on the interaction strength of the $p$-wave contacts. Since the $p$-wave contacts connect the macroscopic properties to the thermodynamic properties of the system, our results would be helpful for further understanding many-body properties of this anisotropic interacting Fermi gas in the normal phase. 
\end{abstract}
\maketitle
\par
\section{Introduction}
One of the most crucial features of ultra cold Fermi gas is a tunability of the interaction strength between Fermi atoms, realized by Feshbach resonance\cite{Chin,Timmermans}. Since, by using the tunable $s$-wave interaction, the BCS-BEC crossover, in which the superfluid properties change from the weak-coupling BCS-like one to the Bose Einstein condensation of strongly binding molecular boson as increasing the interaction strength\cite{Eagles,Leggett,NSR,Melo,OhashiGriffin}, has been realized in the ultracold Fermi gas\cite{Regal,Zwierlein}, strong-coupling effects on various physical quantities have been extensively investigated both experimentally\cite{Stewart2008,Gaebler2010,Sagi2015,Sanner2011,Nascimbene2010,Navon2010,Ku2012,Horikoshi2017} and theoretically\cite{Ohashi2003,Hu2006,Haussmann2007,Fukushima2007,Hu2008,Tsuchiya2009,Chen2009,Hui2010,Tsuchiya2011,Perali2011,Kashimura2012,Tajima2014,Pieter2016}. In addition to the $s$-wave interaction, in cold atom physics, a tunable $p$-wave interaction has already been realized\cite{Regal2,Regal3,Ticknor,Zhang,Schunck,Gunter,Gaebler2,Inaba,Fuchs,Mukaiyama,Maier} and the $p$-wave pair formation has also been detected \cite{Regal3,Zhang,Gaebler2,Inaba,Fuchs}. However the $p$-wave superfuidity has not been realized yet in this system. The main experimental difficulty of this system is a very short life time of the $p$-wave molecules, caused by three-body loss\cite{Castin,Gurarie3}, as well as dipolar relaxations\cite{Gaebler2}. In view of this current experimental status in this research field, it is helpful to start from understanding the many-body properties of this system in the normal phase above the superfluid transition temperature $T_{\rm c}$.     
\par
In this regard, the $p$-wave contacts have recently been measured in a polarized $^{40}$K Fermi gas with the $p$-wave interaction from the weak-coupling regime to the strong-coupling regime\cite{Luciuk}. The concept of the contact was first introduced for a two-component Fermi gas with a contact-type $s$-wave interaction by S. Tan\cite{Tan1,Tan2,Tan3}. In these works, the contact is originally defined as the coefficient of the leading term of the asymptotic behavior of the momentum distribution at large $p$. Surprisingly, it has found that this quantity completely describes how the total energy of the system changes when one increases the interaction strength. Furthermore, the contact is also related not only to the total energy but also to various thermodynamic quantities. These relations between the contact and the thermodynamic properties are known as the Tan's universal relations, which have been experimentally and theoretically verified in the cold atom Fermi gas\cite{Zhang2009,Kuhnle2009,Palastini2010,Schneider2010,Stewart2010,Werner2012,Sagi2012,Hoinka2013}
. In this sense, the contact connects the microscopic properties to the macroscopic (or thermodynamic) properties of the system. 
\par
The concept of the contact has been recently extended to the case with the $p$-wave interaction\cite{Yu,Yoshida,Yoshida2,He,Zhang2,Peng}. One of the most remarkable difference from the $s$-wave contact is that in the $p$-wave case, the contacts consist of 6 components, in principle, because the $p$-wave interaction consist of three components and each component is characterized by two parameters, the $p$-wave scattering volume and the effective range. In Ref. \cite{Luciuk}, by estimating the $p$-wave contacts from two different quantities, the momentum distribution and the spectral weight of the excitations, a part of the universal relations in the $p$-wave interacting Fermi gas has been experimentally proved. However, the observed $p$-wave contact $C_R$ which is related to the effective range cannot be explained by a simple model. Thus, for further understanding of these experimental results, more precise theoretical analysis is necessary. Especially, the strong-coupling and the finite temperature effects might be important because the measurement of the $p$-wave contacts has been done in the normal phase above $T_{\rm c}$ and for a wide range of the interaction strength. In addition, more recently, it has been theoretically pointed out that, in the sub-leading term of the asymptotic form of the momentum distribution, there is an extra term, which are not related to the thermodynamic properties but comes from the center-of-mass motion of the pair molecules\cite{Peng}. Unfortunately, the existence of this extra term was not considered in the experiment. Thus, in order to describe the experimental results, it is necessary to calculate not only the $p$-wave contacts but also the extra term.
\par
In this paper, we theoretically clarify the strong-coupling and the finite temperature effects on the $p$-wave contacts, defined from the adiabatic energy relations, as well as the asymptotic behavior of the large momentum distribution of a one-component Fermi gas with a tunable $p$-wave interaction. Including $p$-wave pairing fluctuations within the strong-coupling theory developed by Nozi\`eres and Schmitt-Rink (NSR)\cite{NSR,Ohashi2005,Inotani2016,Inotani2017_1,Inotani2017_2}, we numerically evaluate the temperature dependence of these quantities in the entire coupling regime above the superfluid transition temperature $T_{\rm c}$. By comparing the $p$-wave contacts with the asymptotic behavior we show that the contribution from the extra terms becomes remarkable in high temperature region. Including the effects of the harmonic trap potential within the local density approximation (LDA)\cite{Pethick}, we calculate these quantities in the trapped system, which are compared with the recent experimental measurement\cite{Luciuk}. 
\par
This paper is organized as follows. In Sec. II, we explain our strong-coupling theory for an ultracold Fermi gas with a $p$-wave interaction. In Sec. III, we show our numerical results on $T_{\rm c}$. Here we clarify how $p$-wave pairing fluctuations affects the $p$-wave contacts in uniform case. In Sec. IV, we determine the temperature dependence of the $p$-wave contacts and the asymptotic behavior of the momentum distribution. We show that the extra terms in the asymptotic form becomes remarkable in high temperature region. In Sec. V, we extend our theory to the trapped system and we directly compare our numerical results with the resent experiment on $^{40}$K Fermi gas\cite{Luciuk}. Throughout this paper, we take $\hbar=k_{\rm B}=1$, and the system volume $V$ is taken to be unity, for simplicity.
\par
\par
\section{Formulation}
\par
We consider a one-component Fermi gas with a tunable $p$-wave interaction $V_p({\bm p},{\bm p}')$, described by a Hamiltonian\cite{Ohashi2005,Inotani2016,Inotani2017_1,Inotani2017_2,Gurarie,Gurarie2,Inotani2,Inotani3,Iskin,Iskin2,Cheng2006},
\begin{equation}
H=\sum_{\bm p} \xi_{\bm p}c_{\bm p}^{\dagger}c_{\bm p}
+\frac{1}{2}\sum_{{\bm p},{\bm p}',{\bm q}} 
V_p({\bm p},{\bm p}')
c_{{\bm p}+{\bm q}/2}^\dagger c_{-{\bm p}+{\bm q}/2}^\dagger
c_{-{\bm p}'+{\bm q}/2}c_{{\bm p}'+{\bm q}/2}.
\label{eq1}
\end{equation}
Here $c_{\bm p}$ is an annihilation operator of a Fermi atom with a kinetic energy $\xi_{\bm p}=p^2/(2m)-\mu$ measured from the chemical potential $\mu$ (where $m$ is the mass of Fermi atom). In Eq. (\ref{eq1}) the $p$-wave interaction $V_p({\bm p},{\bm p}')$ is assumed to have a separable form as
\begin{equation}
V_p({\bm p},{\bm p}')=-\sum_{i=x,y,z} 
\gamma_i({\bm p}) 
U_i 
\gamma_i({\bm {p}'}),
\label{eq2}
\end{equation}
with tunable interaction strength $U_i$ in the $p_i$-wave Cooper channel ($i=x,y,z$). Here $\gamma_i ({\bm p})$ is the basis function of the $p_i$-wave interaction given by 
\begin{equation}
\gamma_i({\bm p})=p_i F^{(i)}_{\rm c}({\bm p}).
\label{eq3}
\end{equation}
In Eq. (\ref{eq3}), to eliminate the well-known ultraviolet divergence, we introduce a component-dependent cutoff function
\begin{equation}
F_c^{(i)} \left( p \right)=
\frac{1}{1+\left( p_c^{(i)} / p \right)^{2n}},
\label{eq4}
\end{equation}
where $p_c^{(i)}$ is a cutoff momentum of the $p_i$-wave interaction. Throughout this paper, we take $n=3$ in Eq. (\ref{eq4}). However, the physical properties are not remarkably affected by the detail of the high-momentum structure of the $p$-wave interaction in the normal phase discussed in this paper\cite{Inotani3}.
\par
In cold atom physics the $p$-wave interaction is conveniently characterized in terms of the scattering volume $v_i$ and the effective range $R_i$, which are respectively related to $U_i$ and $F_c^{(i)} \left( p \right)$ as
\begin{align}
{4\pi v_i \over m}&=
-{U_i \over 3}
{1 \over \displaystyle 1-{U_i \over 3}\sum_{\bm p}
{{\bm p}^2 \over 2\varepsilon_{\bm p}} \left[ F_{\rm c}^{(i)}({\bm p})\right]^2},
\label{eq5}
\\
R_{i}^{-1}&={2\pi \over m^2}
\sum_{\bm p}
{{\bm p}^2 \over 2\varepsilon_{\bm p}^2} \left[ F_{\rm c}^{(i)}({\bm p}) \right]^2.
\label{eq6}
\end{align}
In $^{40}$K Fermi gas, due to the magnetic dipole-dipole interaction between $p$-wave Feshbach molecules, the $p_z$-wave Feshbach resonance splits from the other degenerate $p_x$ and $p_y$-wave Feshbach resonances\cite{Ticknor} where $z$-axis is taken to be parallel to an external magnetic field. As a result, the realized $p$-wave interaction $V_p({\bm p}, {\bm p}')$ essentially has uniaxial anisotropy, leading component-dependence of $v_{i}$ and $R_i$. However, for the effective range $R_i$ the anisotropy, as well as the field-dependence is not remarkable as reported from the experiment on $^{40}$ K Fermi gas\cite{Ticknor}. Thus we simply assume the effective range is channel-independent as $R_x=R_y=R_z=R$ and we take $R=0.02p_{\rm F}^{-1}$ following the experimental situation in Ref. \cite{Luciuk}. In this assumption, the  anisotropy of the $p$-wave interaction is described by the difference of the inverse scattering volume between the stronger $p_z$-wave component and the weaker $p_x$ and $p_y$-component as $v_z^{-1} \geq v_x^{-1}=v_y^{-1}$. Thus we introduce a parameter $\delta v^{-1}=v_z^{-1}-v_x^{-1}=v_z^{-1}-v_y^{-1} > 0$ describing the strength of the anisotropy. We briefly note that in our model we only consider the case near the $p_z$-wave resonance. Although, near the $p_x$ and $p_y$-wave resonances in $^{40}$K Fermi gas, the $p_z$-wave interaction is expected to be weakly repulsive, this situation cannot be described in our model where all $p$-wave interaction is assumed to be attractive. In addition, we mention that the $p$-wave Feshbach resonance has also been observed in ${^6}$Li Fermi gas\cite{Schunck,Gunter,Gaebler2,Inaba,Fuchs,Mukaiyama,Maier}, where, in contrast to $^{40}$K case, the splitting of the $p$-wave Feshbach resonances has not been reported, that means in this system the anisotropy of the $p$-wave interaction is weak compared to $^{40}$K Fermi gas. Thus we also treat $\delta v^{-1}$ as a tunable parameter, and we clarify how the anisotropy affects the $p$-wave contacts. In this scale, the interaction strength is described by only $v_z^{-1}$ for given $\delta v^{-1}$, and the weak-coupling BCS regime and the strong-coupling BEC regime are characterized as $(v_zk_{\rm F}^3)^{-1} < 0$ and $(v_zk_{\rm F}^3)^{-1} > 0$, respectively (where $k_{\rm F}$ is the Fermi momentum).
\par
\begin{figure}
\centerline{\includegraphics[width=8cm]{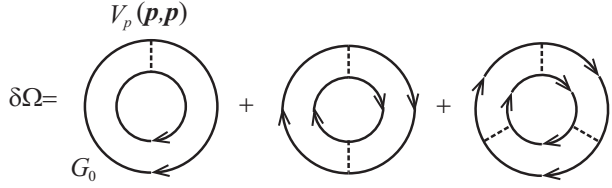}}
\caption{Feynman diagrams describing the strong-coupling correction $\delta \Omega$ to the thermodynamic potential $\Omega$ within the NSR theory. The solid and dashed line describe the bare single-particle Green's function $G_0 \left( {\bm p}, i\omega_m \right)=\left(i\omega_m-\xi_p \right)^{-1}$ and the $p$-wave interaction $V_{p}({\bm p},{\bm p}')$, respectively.}
\label{fig1}
\end{figure}
\par
To investigate strong-coupling and finite temperature effects on the $p$-wave contacts above the $p$-wave superfluid transition temperature $T_{\rm c}$, we include fluctuations in the $p$-wave Cooper channel within a strong-coupling theory developed by Nozi\`eres and Schmitt-Rink (NSR theory)\cite{NSR,Ohashi2005,Inotani2016,Inotani2017_1,Inotani2017_2}. In this approximation, the total thermodynamic potential is given by 
\begin{equation}
\Omega=\Omega_0+\delta\Omega,
\label{thermo}
\end{equation}
where
\begin{equation}
\Omega_0=T \sum_{{\bm p}} \ln \left[1+e^{-\xi_{\bm p}/T} \right]
\label{eq7}
\end{equation}
is the non-interacting part and the strong-coupling correction $\delta \Omega$ is diagramatically shown in Fig. \ref{fig1}. Summing up these diagrams, we obtain 
\begin{equation}
\delta\Omega=T\sum_{{\bm q},\nu_n}
{\rm Tr}\ln \left[-\hat{\Gamma}^{-1}({\bm q}, i\nu_n)\right].
\label{eq8}
\end{equation}
Here $\hat{\Gamma}=\{ \Gamma_{i,j} \}$ ($i,j=x,y,z$) is a 3$\times$3 particle-particle scattering matrix in the $p$-wave Cooper channel given by
\begin{equation}
{\hat \Gamma}({\bm q},i\nu_n)=
{-{\hat U}_p \over 1-{\hat U}_p{\hat \Pi}({\bm q},i\nu_n)},
\label{eq9}
\end{equation}
where, ${\hat U}_P={\rm diag}[U_x,U_y,U_z]$, and
\begin{equation}
\Pi_{i,j}({\bm q},i\nu_n)= 
\frac{1}{2}\sum_{\bm {p}} 
\gamma_i({\bm p}) 
\frac{\tanh \left( \frac{\xi_{\bm{p} + \frac{\bm q}{2}}}{2T} \right)
+\tanh \left( \frac{\xi_{-\bm{p} + \frac{\bm q}{2}}}{2T} \right)}
{\xi_{\bm{p} + \frac{\bm q}{2}}+\xi_{-\bm{p} + \frac{\bm q}{2}}-i\nu_n}
\gamma_j({\bm p})
\label{eq10}
\end{equation}
is the $3\times 3$-matrix lowest-order pair-correlation function.
\par
Before moving on to the calculation of the $p$-wave contacts, we briefly summarize the Tan's contact in an $s$-wave interacting Fermi gas. In this case, the Tan's contact $C_s$ is originally defined from the asymptotic behavior of the large-momentum distribution $n_{\bm p}$ as
\begin{equation}
n_{\bm p} \to \frac{C_s}{p^4},~~~~~~(s{\rm -wave})
\end{equation}
where high momentum limit $k_{\rm F} \ll p$ is taken. $C_s$ is also proportional to the derivative of the total energy of the system with respect to the inverse $s$-wave scattering length $a_s^{-1}$ as 
\begin{equation}
C_s=-4 \pi m\frac{\partial E}{\partial a_s^{-1}}.~~~~~~(s{\rm -wave})
\label{aer}
\end{equation}
This equation is known as the adiabatic energy relation. 
\par
In the $p$-wave case, due to the momentum dependence of the $p$-wave interaction $V_p({\bm p},{\bm p}')$, the asymptotic form of the momentum distribution consists of two terms as
\begin{align}
n_{\bm p} \to \frac{C_2 \left(\hat{\bm p} \right)}{p^2}+ \frac{C_4 \left(\hat{\bm p} \right)}{p^4},~~~~~~(p{\rm -wave})
\label{asymp}
\end{align}
where $\hat{\bm p}={\bm p}/|{\bm p}|$. On the other hand, the adiabatic energy relation Eq. (\ref{aer}) is straightforwardly extended to the $p$-wave case as
\begin{align}
C_v^{(i)}&= -2m \left( \frac{\partial \Omega}{\partial v_i^{-1}} \right)_{T,\mu},
\label{eq13}
\\
C_R^{(i)}&= - 2m \left( \frac{\partial \Omega}{\partial R_i^{-1}} \right)_{T,\mu}.
\label{eq14}
\end{align}
We note that since the $p$-wave interaction is characterized by 6 parameters ($v_i$ and $R_i$ for $i=x,y,z$), the $p$-wave contacts also consist of 6 components, in principle. For the leading term in Eq. (\ref{asymp}) ($\propto p^{-2}$), $C_v^{(i)}$ defined from Eq. (\ref{eq13}) completely capture the coefficient $C_2$ as $C_2 (\hat{\bm p}) \propto \sum_i C_v^{(i)} \hat{p}_i^2$. However, as first mentioned in \cite{Peng}, $C_4$ in the sub-leading term cannot be written by only $C_R^{(i)}$ and the extra terms appear. These extra terms also exist within our theoretical framework. To clearly see this, although in the original Tan's work\cite{Tan1,Tan2,Tan3} the contact is defined from the momentum distribution, here, we first define the $p$-wave contacts $C_v^{(i)}$ and $C_R^{(i)}$ from the adiabatic energy relations Eq. (\ref{eq13}) and Eq. (\ref{eq14}), respectively. Substituting the thermodynamic potential $\Omega$ Eq. (\ref{thermo}) into Eqs. (\ref{eq13}) and (\ref{eq14}), and using Eqs. (\ref{eq5}) and (\ref{eq6}) to eliminate $U_i$ and $F_c^{(i)}$, we obtain the expressions for the $p$-wave contacts $C_v^{(i)}$ and $C_R^{(i)}$ within the NSR theory as
\begin{align}
C_v^{(i)}&=
-\frac{m^2}{6 \pi \beta} \sum_{{\bm q}, \nu_n} \Gamma_{i,i} \left({\bm q}, i\nu_n \right),
\label{eq15}
\\
C_R^{(i)}&=
\frac{m^3}{6 \pi \beta} \sum_{{\bm q}, \nu_n} 
\left( \varepsilon_q^{\rm B} - i \nu_n -2\mu \right) 
\Gamma_{i,i} \left({\bm q}, i\nu_n \right),
\label{eq16}
\end{align}
where $\varepsilon_q^{\rm B} = q^2/(4m)$ is a kinetic energy of a pair molecule with center-of-mass momentum ${\bm q}$ and $\nu_n=2 \pi n T$ is boson Matsubara frequency.
On the other hand, the momentum distribution $n_{\bm p}$ is conveniently calculated from the single-particle Green's function $G({\bm p}, i\omega_m)$ as   
\begin{equation}
n_{\bm p}=\frac{1}{\beta}\sum_{\omega_m} G\left( {\bm p} , i \omega_m\right),
\label{eq17}
\end{equation}
where $\omega_m=(2m+1)\pi T$ is fermion Matsubara frequency.
$G\left( {\bm p} , i \omega_m\right)$ being consistent with the thermodynamic potential Eq. (\ref{thermo}) is given by\cite{Inotani2016,Inotani2017_1,Inotani2017_2}
\begin{equation}
G\left( {\bm p} , i \omega_m\right) 
=
G_0\left( {\bm p} , i \omega_m\right)
+
G_0\left( {\bm p} , i \omega_m\right)
\Sigma\left( {\bm p} , i \omega_m\right)
G_0\left( {\bm p} , i \omega_m\right),
\label{eq18}
\end{equation}
where
\begin{equation} 
\Sigma\left( {\bm p} , i \omega_m\right)
= \frac{2}{\beta} \sum_{{\bm q}, \nu_n}
\sum_{i,j=x,y,z}
\gamma_i \left( {\bm p} - \frac{\bm {q}}{2} \right)
\Gamma_{i,j} \left( {\bm q}, i\nu_n \right)
\gamma_j \left( {\bm p} - \frac{\bm {q}}{2} \right)
G_0\left( -{\bm p}+{\bm q} , -i \omega_m + i \nu_n \right)
\label{eq19}
\end{equation}
is the strong-coupling self-energy correction. In the large momentum region $p_{\rm F} \ll p \ll R^{-1}$, $n_{\bm p}$ becomes
\begin{equation}
n_{\bm p} \to 
12 \pi \sum_i C_v^{(i)} \frac{\hat{p}_i^2}{p^2} 
+\left [24 \pi \sum_i \left( C_R^{(i)} +\eta_i \right) \hat{p}_i^2
+6 \pi \zeta
+48 \pi \sum_{i,j} \kappa_{i,j} \hat{p}_i^2\hat{p}_j^2 \right] \frac{1}{p^4},
\label{eq20}
\end{equation}
where $\hat{p}_i=p_i/|{\bm p}|$. Indeed, we find that the extra terms $\zeta$, $\eta_i$ and $\kappa_{ij}$ appear in the sub-leading term in Eq. (\ref{eq20}). The coefficients of these extra terms, $\zeta$, $\eta_i$ and $\kappa_{ij}$ are given by
\begin{align}
\zeta&=-\frac{m^2}{12 \pi \beta} \sum_{{\bm q}, \nu_n} \sum_{i,j}
q_iq_j \Gamma_{i,j}(\bm{q},i\nu_n),
\label{eq21}
\\
\eta_i&=\frac{m^2}{12 \pi \beta} \sum_{{\bm q}, \nu_n}  
\left[
\left( \frac{q^2}{2} 
+2q_i^2 \right)
\Gamma_{i,i}(\bm{q},i\nu_n)
+q_i\sum_{j \neq i}  q_j \left( \Gamma_{i,j}(\bm{q},i\nu_n) + \Gamma_{j,i}(\bm{q},i\nu_n) \right)
\right], 
\label{eq22}
\\
\kappa_{i,j}&=
\left\{
\begin{array}{l}
\displaystyle
- \frac{m^2}{8 \pi \beta} \sum_{{\bm q},\nu_n}
q_i^2 \Gamma_{i,i} \left( {\bm q}, i\nu_n \right),~~~~~~(i=j)
\\
\displaystyle
- \frac{m^2}{16 \pi \beta} \sum_{{\bm q},\nu_n}
\left[
q_i^2 \Gamma_{j,j} \left( {\bm q}, i\nu_n \right)+q_j^2\Gamma_{i,i} \left( {\bm q}, i\nu_n \right)
+2q_iq_j \left( \Gamma_{i,j} \left( {\bm q}, i\nu_n \right)+\Gamma_{j,i} \left( {\bm q}, i\nu_n \right) \right)
\right].~~~~~(i \neq j)
\end{array}
\right.
\nonumber
\\
\label{eq23}
\end{align}
From Eqs. (\ref{eq21})-(\ref{eq23}), we find that $\zeta$, $\eta$, and $\kappa$ result from the center-of-mass motion of the pairs as discussed in Ref. (Note that $\Gamma_{i,j}\left({\bm q} , i\nu_n \right)$ physically describes the pair propagation.) At $T=0$, since the thermal excitations of pairs are completely suppressed, the extra terms are expected to be negligible. However, in the normal phase above $T_{\rm c}$, the preformed Cooper pairs with finite momentum ${\bm q} \neq 0$ might exist. As we will show in Sec. \ref{t_dependence}, in fact, the contribution from $\zeta$, $\eta_i$ and $\kappa_{i,j}$ to the sub-leading term of $n_{\bm p}$ becomes more dominant than $C_R^{(i)}$ in a high temperature regime. We will also discuss in Sec. \ref{trap} how strong the effects of them are in the experimental situation in Ref.\cite{Luciuk}. We briefly note that, using the rotational symmetry around $z$-axis of $V_p({\bm p}, {\bm p}')$ in Eq. (\ref{eq1}), we obtain $C_v^{(x)}=C_v^{(y)}$, $C_R^{(x)}=C_R^{(y)}$, $\eta_x=\eta_y$, $\kappa_{x,z}=\kappa_{y,z}$, and $\kappa_{x,x}=\kappa_{y,y}=\kappa_{x,y}=\kappa_{y,x}$ 
\par
To calculate $C_v^{(i)}$ and $C_R^{(i)}$, as well as $\zeta$, $\eta_i$ and $\kappa_{ij}$, in the normal phase, we need to determine the suprefluid transition temperature $T_{\rm c}$ and the chemical potential $\mu \left( T \right)$. In the present uniaxially anisotropic case ($U_z \geq U_x = U_y$), the $p$-wave superfluid transition from the normal phase always occurs in the stronger $p_z$-wave Cooper channel. Thus the equation for $T_{\rm c}$ is obtained from the Thouless criterion in the $p_z$-wave channel $\Gamma^{-1}_{zz}(0,0)=0$\cite{Inotani2016,Inotani2017_1,Inotani2017_2,Inotani3}, leading
\begin{equation}
1=U_z \sum_{\bm p} \frac{\gamma_z^2\left( {\bm p} \right) }{2\xi_p} \tanh \left( \frac{\xi_p}{2T} \right).
\label{eq24}
\end{equation}
By solving Eq. (\ref{eq24}) together with the particle number equation obtained from the thermodynamic potential $\Omega$ as
\begin{align}
N&=-\left( \frac{\partial \Omega}{\partial \mu} \right)_{T}
\nonumber
\\
&=
\sum_{\bm p} n_{\rm F}(\xi_{\bm p})
+
T\sum_{{\bm q},\nu_n}
{\rm Tr}
\left[\hat{\Gamma}({\bm q},i\nu_n)
\left(\frac{\partial {\hat \Pi}({\bm q},i\nu_n)}{\partial \mu} \right)_T
\right],
\nonumber
\\
\label{eq25}
\end{align}
we determine $T_{\rm c}$ and $\mu \left(T=T_{\rm c}\right)$. After that, the chemical potential above $T_{\rm c}$ is simply obtained by solving only Eq. (\ref{eq25}) for given $T$ ($> T_{\rm c}$), $v_z^{-1}$ and $\delta v^{-1}$.
\par
For the comparison with the recent experiment, here, we summarize how to measure the $p$-wave contacts from the momentum distribution. Experimentally, the momentum distribution is observed by taking absorption image after time-of-flight with an imaging beam. The imaging beam used in Ref. \cite{Luciuk} propagates along $z$-direction parallel to the Feshbach magnetic field. Then one obtains the 2-dimensional momentum distribution $\bar{n}(p_x,p_y)$ which is the averaged $n_{\rm p}$ over the $p_z$ direction as $\bar{n}\left(p_x,p_y \right) = \int \frac{dp_z}{2\pi} n_{\bm p}$. When we ignore the extra terms from Eq. (\ref{eq20}), we obtain a formula 
\begin{equation}
\bar{n}\left(p_x,p_y \right)  \to 
\frac{3 \pi  \bar{C}_v}{\rho} 
+ 
\frac{3 \pi  \bar{C}_R}
{2 \rho^3},
\label{eq26}
\end{equation}
where $\rho=\sqrt{p_x^2+p_y^2}$ and
\begin{align}
\bar{C}_v&=C_v^{(z)}+\frac{1}{2}\sum_{i=x,y} C_v^{(i)},
\label{eq27}
\\
\bar{C}_R&=C_R^{(z)}+\frac{3}{2}\sum_{i=x,y} C_R^{(i)}.
\label{eq28}
\end{align}
We briefly note that the coefficients $1/2$ in Eq. (\ref{eq27}) and $3/2$ in Eq. (\ref{eq28}) for the $p_x$- and $p_y$-component of the $p$-wave contacts are due to the anisotropy of the momentum distribution. They also assume that $C_v^{x}=C_v^{y}$ and $C_R^{x}=C_R^{y}$ can be negligible near the $p_z$-wave resonance field because the $p_z$-wave resonance and the $p_x$ and $p_y$-wave resonance are well-separated in $^{40}$K Fermi gas. Finally, they obtain $C_v^{(z)}$ and $C_R^{(z)}$ near the $p_z$-wave resonance by fitting.
\par
The effects from the extra terms do not considered in Eq. (\ref{eq26}), but they actually modify Eq. (\ref{eq26}) as
\begin{equation}
\bar{n}(\rho)
\to 
\frac{3 \pi  \bar{C}_v}{\rho} 
+ 
\frac{3 \pi  \left( \bar{C}_R+\delta \right)}
{2 \rho^3}.
\label{eq29}
\end{equation}
where 
\begin{equation}
\delta=\zeta+\eta_z+\frac{3}{2} \sum_{i=x,y} \eta_{i}
+\kappa_{z,z}+\sum_{i=x,j} \left(\kappa_{i,z}+ \kappa_{z,i}\right)
+\frac{5}{3}\sum_{i,j=x,y} \kappa_{i,j}
\label{eq30}
\end{equation}
is the correction, coming from the extra terms, to the coefficient of the sub-leading term ($\propto \rho^{-3}$) of $\bar{n}\left(p_x,p_y \right)$. Thus, the experimental result for $C_{R}^{(z)}$ should be compared not with $\bar{C}_R$ but with $\bar{C}_R+\delta$
\par
Since the measurement has been done under the trap potential for confining the atoms, the measured $\bar{n}(p_x,p_y)$ is also spatially averaged over the gas cloud. Thus we also need to consider the effects of the trap potential to directly compare with the experiment. Using the local density approximation (LDA)\cite{Pethick}, this is simply achieved by replacing $\mu$ by the position dependent chemical potential $\mu(r)=\mu-V(r)$ where $V(r)=m\omega^2 r^2/2$ is a harmonic potential at the position $r$ measured from the trap center ($r=0$) with the trap frequency $\omega$. Then, Eqs. (\ref{thermo})-({\ref{eq10}}) become position dependent. From the total thermodynamic potential $\int d{\bm r} \Omega(r)$, the LDA particle number equation is given by 
\begin{align}
N&=- \int d{\bm r} \left( \frac{\partial \Omega(r)}{\partial \mu(r)} \right)_{T},
\label{eq31}
\end{align}
and the equation for $T_{\rm c}$ is derived from the Thouless criterion at the trap center $r=0$. $C_v^{(i)}$ and $C_R^{(i)}$, as well as $\zeta$, $\eta_i$ and $\kappa_{i,j}$ in the trapped system are obtained by replacing $\mu$ by $\mu(r)$ and carrying out the ${\bm r}$ integral. For example,
\begin{align}
C_v^{(i)}&= -2m \int d{\rm r} \left( \frac{\partial \Omega(r)}{\partial v_i^{-1}} \right)_{T,\mu},
\label{eq32}
\\
C_R^{(i)}&= - 2m \int d{\rm r}  \left( \frac{\partial \Omega(r)}{\partial R_i^{-1}} \right)_{T,\mu}.
\label{eq33}
\end{align}
\par
In Sec. III and IV, we first focus on the $p$-wave contacts and the extra terms in the uniform system and we will discuss how strong pairing fluctuations and finite temperature affect these quantities. After that, including the effects of the harmonic trap potential, we directly compare our results with the recent measurement in Sec. V.
\section{$p$-wave contacts at the superfluid transition temperature in uniform system}
\label{tc}
\par
\begin{figure}
\centerline{\includegraphics[width=7cm]{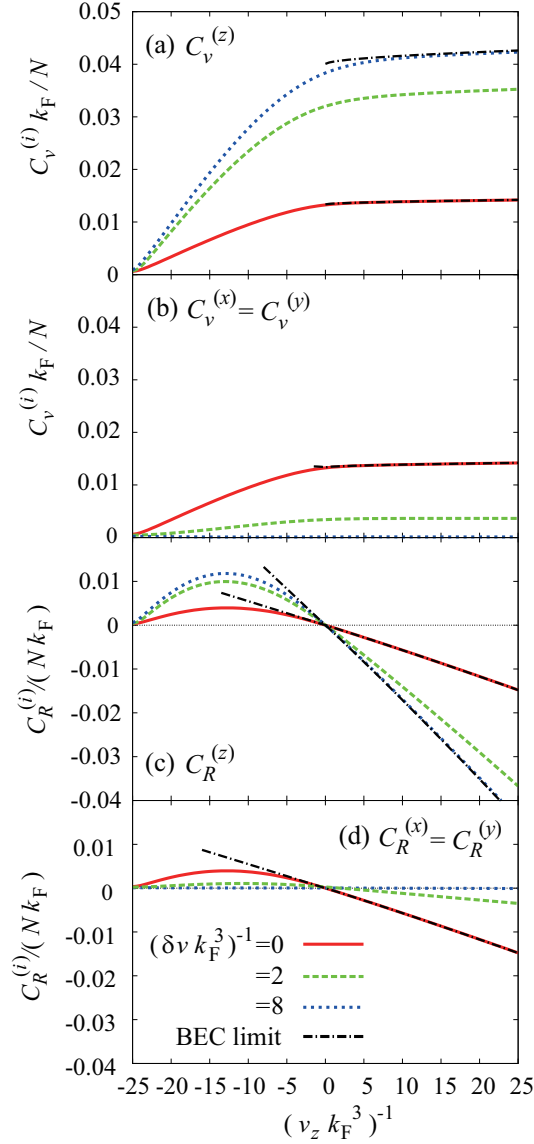}}
\caption{(Color online) calculated $p$-wave contacts $C_v^{(i)}$ and $C_R^{(i)}$ ($i=x,y,z$) at the superfluid transition temperature $T_{\rm c}$ as a function of the strength of the $p_z$-wave interaction $(v_z k_{\rm F}^3)^{-1}$ for three typical anisotropic parameter [$(\delta v k_{\rm F}^3)^{-1}=0,2,8$]. The chain lines in each panel shows $C_v^{(i)}$ and $C_R^{(i)}$ in the BEC limit in the isotropic case and the anisotropic limit.}
\label{fig2}
\end{figure}
\par
Figure \ref{fig2} shows the $p$-wave contacts (a), (b) $C_
v^{(i)}$ and (c), (d) $C_R^{(i)}$ for $i=x,y,z$ in a uniform one-component Fermi gas with a tunable $p$-wave interaction with a uniaxial anisotropy at $T=T_{\rm c}$ as a function of the interaction strength $(v_z k_{\rm F}^3)^{-1}$ for $(\delta v k_{\rm F}^3)^{-1}=0,2,8$. As shown in Fig. \ref{fig2} (a) and (b), $C_v^{(i)}$ monotonically increases as a function of $(v_zk_{\rm F}^3)^{-1}$ from the weak-coupling regime to the strong-coupling regime. On the other hand, from Fig. \ref{fig2} (c) and (d), we find that although $C_R^{(i)}$ first increases in the weak-coupling regime [$(v_zk_{\rm F}^3)^{-1} \lesssim 12$], as further increasing $(v_zk_{\rm F}^3)^{-1}$, $C_{R}^{(i)}$ decreases and vanishes around $(v_zk_{\rm F}^3)^{-1} = 0$. Eventually, $C_{R}^{(i)}$ becomes negative in the strong-coupling regime $(v_zk_{\rm F}^3)^{-1} > 0$. 
\par
To explain the different interaction-dependence between $C_v^{(i)}$ and $C_R^{(i)}$, we start from the strong-coupling BEC limit. Noting that, in this limit, $\hat{\Gamma}({\bm q} , i\nu_n)$ is reduced to the single-particle Bose Green's function as
\begin{equation}
\Gamma_{i,j}({\bm q} , i\nu_n)=\frac{1}{\alpha\left( \mu \right) \left[ i\nu_n -\left(\varepsilon_q^B-\mu_B^{(i)} \right) \right]}\delta_{i,j}.
\label{eq34}
\end{equation}
Here
\begin{equation}
\alpha\left( \mu \right) =\sum_{\bm p} \frac{F_{\rm c}^2\left( p \right)}{12\xi_p} \simeq \frac{m^2}{12\pi} \left( R^{-1} -\frac{3}{2}\sqrt{2m|\mu|} \right),
\label{eq35}
\end{equation}
and
\begin{align}
\mu_B^{(z)}&=0, 
\label{eq36}
\\
\mu_B^{(x)}&=\mu_B^{(y)}=-\frac{m\delta v^{-1}}{12\pi \alpha\left( \mu \right)},
\label{eq37}
\end{align}
is the chemical potential of the $p_i$-wave molecular Boson. In deriving Eq. (\ref{eq34}), we use the fact that $|\mu| \gg T_{\rm c}$ in this limit. Substituting Eq. (\ref{eq34}) into Eqs. (\ref{eq15}) and (\ref{eq16}), we obtain the strong-coupling expressions for $C_v^{(i)}$ and $C_R^{(i)}$ as
\begin{align}
C_v^{(i)} &\simeq \frac{2}{R^{-1}-3\sqrt{2m|\mu|}/2} N_{\rm B}^{(i)} \simeq -2m \frac{\partial E_{\rm{bind}}^{(i)}}{\partial v_i^{-1}} N_{\rm B}^{(i)},
\label{eq38}
\\
C_R^{(i)}  &\simeq \frac{4m\mu N}{R^{-1}-3\sqrt{2m|\mu|}/2} N_{\rm B}^{(i)} \simeq -2m\frac{\partial E_{\rm{bind}}^{(i)}}{\partial R_i^{-1}} N_{\rm B}^{(i)},
\label{eq39}
\end{align}
where $N_{\rm B}^{(i)} = \sum_{\bm q} n_{\rm B} \left( \varepsilon_q^{\rm B}-\mu_{\rm B}^{(i)} \right)$ and $E_{\rm {bind}}^{(i)}=-R/(mv_i)$ are, respectively, the number and the two-body binding energy of the $p_i$-wave molecular Boson [$n_{\rm B} (\varepsilon)$ is the Bose distribution function]. In figure \ref{fig2}, Eqs. (\ref{eq38}) and (\ref{eq39}) in the isotropic limit [$(\delta vk_{\rm F}^3)^{-1}=0$] and in the anisotropic limit [$(\delta vk_{\rm F}^3)^{-1}=\infty$] are also shown (see the chain lines in each panel) and we find these expressions well-describe the behavior of $C_v^{(i)}$ and $C_R^{(i)}$ in the strong-coupling regime [$(v_zk_{\rm F}^3)^{-1} \gesim 0$].  
\par
Physical meaning of these strong-coupling expressions can be understood as follows. In the BEC limit, because all Fermi atoms form molecular Boson due to the strong attractive interaction, the total energy of the system is given by $\sum_i E_{\rm {bind}}^{(i)} N_{\rm {B}}^{(i)}$. Furthermore, the number of the molecules does not change even when $v_i$ and $R_i$ are varied as far as $|E_{\rm{bind}}| \ll T$. Thus, the last expression of Eqs. (\ref{eq38}) and (\ref{eq39}) are, indeed, proportional to the derivative of the total energy with respect to $v_i^{-1}$ and $R_i^{-1}$, respectively. That is consistent with the adiabatic energy relations. 
\par
From the relation between $v_i$, $R_i$ and $U_i$, $p_c^{(i)}$ [Eqs. (\ref{eq5}) and (\ref{eq6})], when one increases $v_i^{-1}$ with fixing $R$, only $U_i$ increases, but $p_c^{(i)}$ does not change. Thus the increase of $v_i^{-1}$ simply means the increase of the coupling constant $U_i$. Thus $E_{\rm{bind}}<0$ is an decreasing function of $v_i^{-1}$, leading the positive value of $C_v^{(i)}$ in the whole interaction region. On the other hand, when one increases $R^{-1}$ with fixing $v_i^{-1}$, we find that $p_c^{(i)}$ increases, but $U_i$ decreases at the same time to fix $v_i^{-1}$ [see Eq. (\ref{eq5})]. Thus there is a competition between the increase of $p_c^{(i)}$ which enhances the interaction and the decrease of $U_i$ which suppresses the interaction with increasing $R^{-1}$. In the BEC limit, the latter is more dominant than the former. As a result, $C_{R}^{(i)}$ has negative value in this regime.
\par
To consider the binding energy $E_{\rm {bind}}^{(i)}$ of the $p_i$-wave pair state in the weak-coupling BCS regime, we need to include the effects of the Fermi surface, which support the pairing phenomenon. Here, we estimate, $E_{\rm {bind}}^{(i)}$ in the BCS regime by applying the Cooper problem to the $p$-wave case, where the equation for $E_{\rm {bind}}^{(i)}$ is given as
\begin{equation}
\frac{1}{U_i}=\sum_{|{\bm p}| \ge p_{\rm F}} \frac{\gamma_i^2\left( {\bm p} \right)}{2\left( \varepsilon_p-\varepsilon_{\rm F} \right)-E_{\rm{bind}}^{(i)}}.
\end{equation}
Using $\varepsilon_{\rm F} \gg E_{\rm{bind}}^{(i)}$ in the weak-coupling limit, we obtain
\begin{equation}
E_{\rm{bind}}^{(i)}=-8\varepsilon_{\rm F} 
\exp 
\left[ 
\pi
\left( 
\frac{1}{k_{\rm F} R}
+\frac{1}{k_{\rm F}^3 v_i} 
\right)
-\frac{8}{3}
\right].
\end{equation}
We find that, in contrast to the strong-coupling regime, $E_{\rm{bind}}^{(i)}$ is a decreasing function of not only $v_i^{-1}$ but also $R^{-1}$. That means the increase of $p_c^{(i)}$ is more significant for $E_{\rm{bind}}^{(i)}$ than the decreases of $U_i$ with increasing $R^{-1}$ and $C_R^{(i)}$ becomes positive. We briefly comment on the effects of the anisotropy. As shown in Fig. \ref{fig2}, $C_v^{(z)}$ and $C_R^{(z)}$ become remarkable when the anisotropy becomes strong, as expected. However, the $(v_zk_{\rm F}^3)^{-1}$-dependence of $C_v^{(i)}$ and $C_R^{(i)}$ are not significantly affected by the anisotropy.
\par
\begin{figure}
\centerline{\includegraphics[width=8cm]{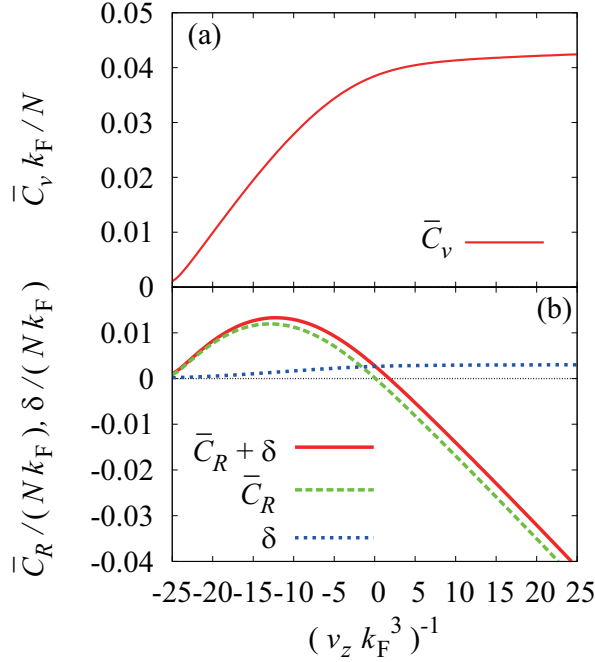}}
\caption{(Color online) coefficient of (a) the leading term $\bar{C}_v$ and (b) the sub-leading term $\bar{C}_R+\delta$ of the asymptotic form of $\bar{n}(p_x,p_y)$ at $T_{\rm c}$ and $(\delta v k_{\rm F}^3)^{-1}=8$, as a function of the interaction strength $(v_z k_{\rm F}^3)^{-1}$. The dashed line and the dotted line in panel (b) show $\bar{C}_R$ and $\delta$, respectively.}
\label{fig3}
\end{figure}
\par
The coefficients of the leading term ($\propto \rho^{-1}$) $\bar{C}_v$ and the sub-leading term ($\propto \rho^{-3}$) $\bar{C}_R + \delta$ of $\bar{n}\left(p_x, p_y \right)$ [Eq. (\ref{eq29})] at $T_{\rm c}$ and $(\delta vk_{\rm F}^3)^{-1}=8$ are shown in Fig. \ref{fig3}. The interaction dependence of $\bar{C}_{v}$ and $\bar{C}_R+\delta$ seems to be similar to one of $C_v^{(i)}$ and $C_R^{(i)}$ shown in Fig. \ref{fig2}. We also note that, at $T=T_{\rm c}$, the contribution from the extra terms $\delta$ does not remarkably affects the results over the entire interaction region [see the dotted line in Fig. \ref{fig3} (b)]. This means the thermal excitations of the pair molecules are sufficiently suppressed, at least at $T_{\rm c}$. 
\section{Temperature dependence of the $p$-wave contacts in uniform system}
\label{t_dependence}
\par
\begin{figure}
\centerline{\includegraphics[width=8cm]{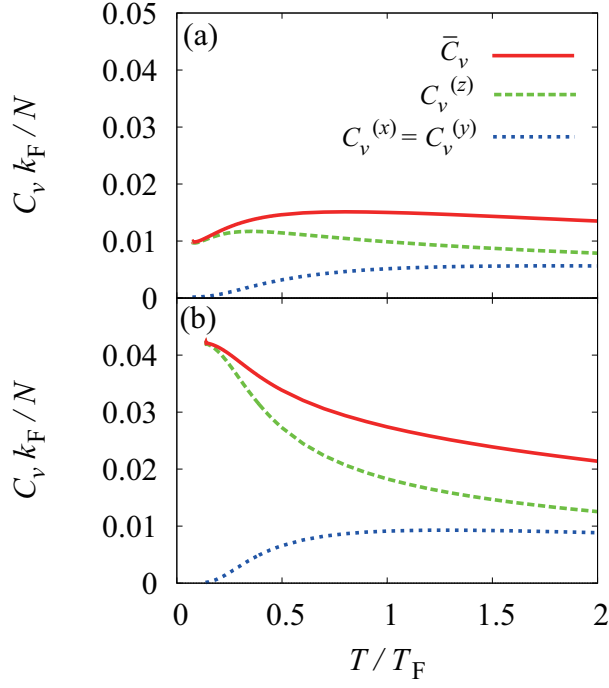}}
\caption{(Color online) Temperature dependence of the $p$-wave contacts related to the scattering volume $\bar{C}_v$ in Eq. (\ref{eq27}), $C_v^{(z)}$ and $C_v^{(x)}=C_v^{(y)}$ in (a) the weak-coupling regime $(v_zk_{\rm F}^3)^{-1}=-20$ and in (b) the strong-coupling regime $(v_zk_{\rm F}^3)^{-1}=-20$. In this figure we take $(\delta v k_{\rm F}^3)^{-1}=8$.}
\label{fig4}
\end{figure}
\par
\begin{figure}
\centerline{\includegraphics[width=8cm]{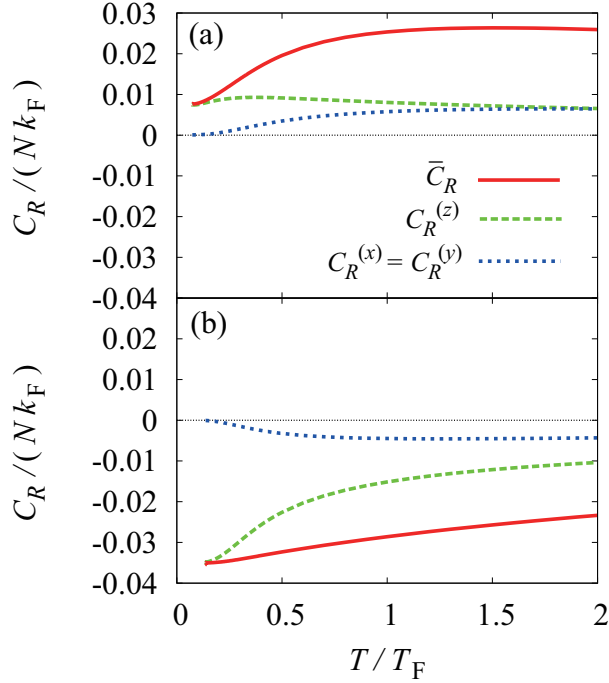}}
\caption{(Color online) Temperature dependence of the $p$-wave contacts related to the effective range $\bar{C}_R$ in Eq. (\ref{eq28}), $C_R^{(z)}$ and $C_R^{(x)}=C_R^{(y)}$ in (a) the weak-coupling regime $(v_zk_{\rm F}^3)^{-1}=-20$ and in (b) the strong-coupling regime $(v_zk_{\rm F}^3)^{-1}=20$. In this figure we take $(\delta v k_{\rm F}^3)^{-1}=8$. }
\label{fig5}
\end{figure}
\begin{figure}
\centerline{\includegraphics[width=8cm]{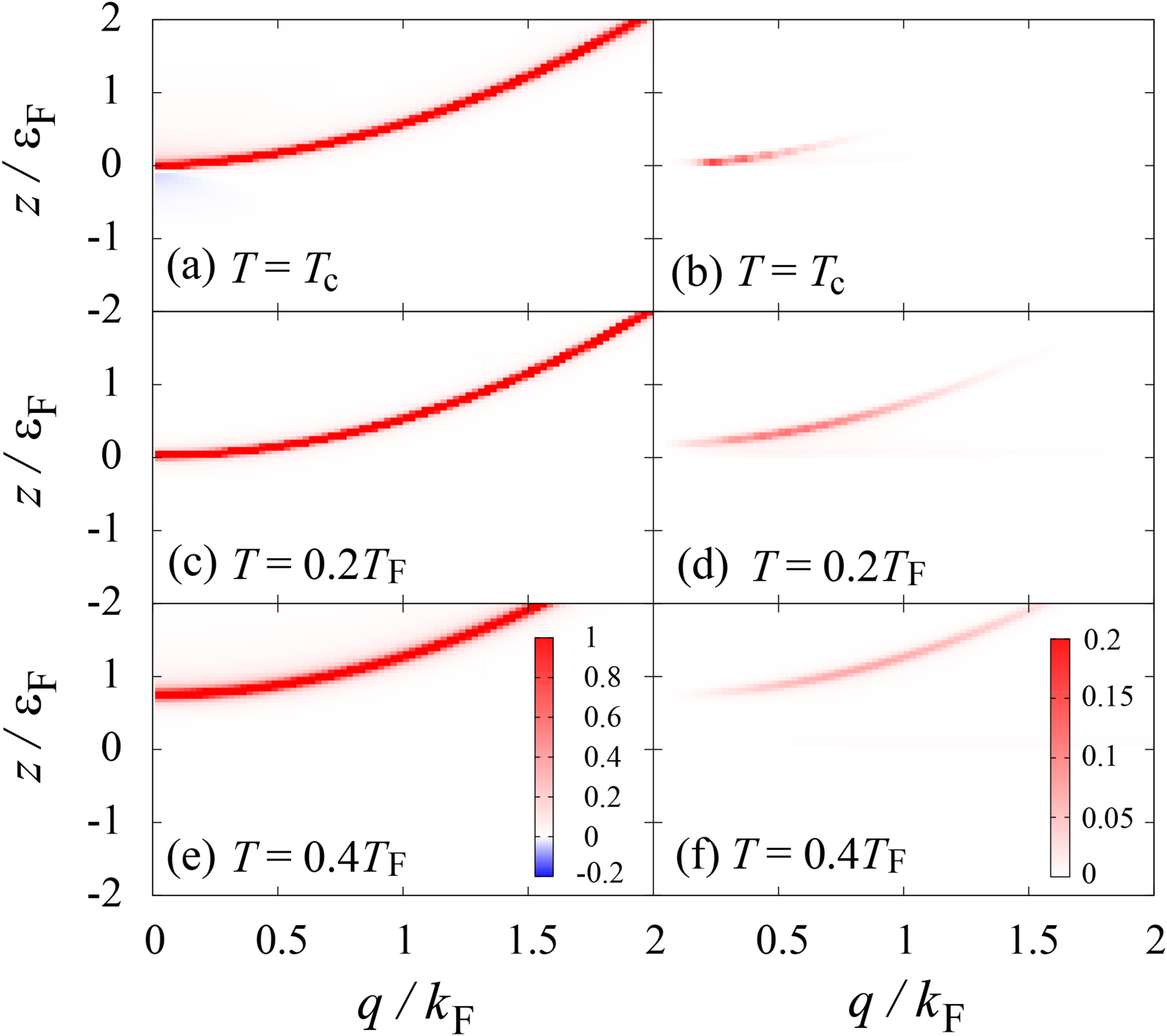}}
\caption{(Color online) (a), (c), (e) calculated ${\rm{Im}} \left[ \Gamma_{z,z}({\bm q},i\nu_n \to z+i\delta) \right]$ and (b), (d), (f) $q^2 n_{\rm B}(z){\rm{Im}} \left[ \Gamma_{z,z}({\bm q},i\nu_n \to z+i\delta) \right]$ in the weak-coupling regime $(v_zk_{\rm F}^3)^{-1}=-20$. In this figure we take ${\bm q}=(0,0,q)$ and $(\delta v k_{\rm F}^3)^{-1}=8$}
\label{fig6}
\end{figure}

Figure \ref{fig4} and \ref{fig5} show the temperature dependence of the $p$-wave contacts $C_v^{(i)}$, $\bar{C}_v$ (in Fig. \ref{fig4}) and $C_R^{(i)}$, $\bar{C}_R$ (in Fig. \ref{fig5}) in the weak-coupling regime $(v_zk_{\rm F}^3)^{-1}=-20$ [Fig. \ref{fig4} (a) and \ref{fig5} (a)] and in the strong-coupling region $(v_zk_{\rm F}^3)^{-1}=20$ [Fig. \ref{fig4}(b) and \ref{fig5}(b)] with $(\delta v k_{\rm F}^3)^{-1}=8$. For $\bar{C}_v$ and $\bar{C}_R$, while in the strong-coupling regime, both $\bar{C}_v$ and $\bar{C}_R$ gradually vanish as increasing $T$, in the weak-coupling regime, we find a non-monotonic temperature dependence of $\bar{C}_v$ and $\bar{C}_R$. The key to understand this non-monotonic behavior is a competition between two finite temperature effects on the $p$-wave contacts. To more clearly see this, we rewrite Eqs. (\ref{eq15}) and (\ref{eq16}) by using the spectral representation as
\begin{align}
C_v^{(i)}&=
-\frac{m^2}{6 \pi^2} \sum_{{\bm q}} \int dz 
{\rm {Im}} \left[
\Gamma_{i,i} \left({\bm q}, i\nu_n \to z+i\delta \right) 
\right]
n_{\rm B} (z)
\label{eq42}
\\
C_R^{(i)}&=
\frac{m^3}{6 \pi^2} \sum_{{\bm q}} \int dz  
\left( \varepsilon_q^{\rm B} - z -2\mu \right) 
{\rm {Im}} \left[
\Gamma_{i,i} \left({\bm q}, i\nu_n \to z+i\delta \right)
\right]
n_{\rm B}(z)
\label{eq43}
\end{align}
Here, $\Gamma_{i,i} \left({\bm q}, i\nu_n \to z+i\delta \right)$ is the analytic continued particle-particle scattering matrix, of which the imaginary part physically describes the spectrum of the collective mode. One of the important finite temperature effects is the change of the structure of ${\rm{Im}} \left[ \Gamma_{i,i} \left({\bm q}, i\nu_n \to z+i\delta \right) \right]$ with increasing $T$. In Fig. \ref{fig6}(a), (c), and (e), the temperature dependence of ${\rm{Im}} \left[ \Gamma_{z,z} \left({\bm q}, i\nu_n \to z+i\delta \right) \right]$ in the weak-coupling regime is shown. At $T=T_{\rm c}$, because of the Thouless criterion Eq. (\ref{eq24}) the collective excitation is gapless. However, as increasing $T$ from $T_{\rm c}$, the peak structure in ${\rm{Im}} \left[ \Gamma_{z,z} \left({\bm q}, i\nu_n \to z+i\delta \right) \right]$ is gradually shifted to higher energy region, that suppresses $\bar{C}_v$ and $\bar{C}_R$. The other finite temperature effect is included $n_{\rm B}(z)$ in Eqs. (\ref{eq42}) and (\ref{eq43}), which describes the occupancy of the collective excitations with energy $z$. 
As increasing $T$, since the thermal excitations become remarkable and the collective mode with high energy ($z \lesssim T$) starts to be occupied, leading the enhancement of $\bar{C}_v$ and $\bar{C}_R$. Fig. \ref{fig6} (b), (d) and (f) show
$q^2{\rm{Im}} \left[ \Gamma_{z,z} \left({\bm q}, i\nu_n \to z+i\delta \right) \right] n_{\rm B} (z)$.
From this figure, we find that the latter effect is more remarkable than the former in the weak-coupling regime near $T_{\rm c}$. 
Thus $\bar{C}_v$ and $\bar{C}_R$ 
first increase with increasing $T$ from $T_{\rm c}$. 
Of course, in the high temperature limit, the effects of the interaction become negligible and $\bar{C}_v$ and $\bar{C}_R$ vanish as expected.
\par
\par
\begin{figure}
\centerline{\includegraphics[width=8cm]{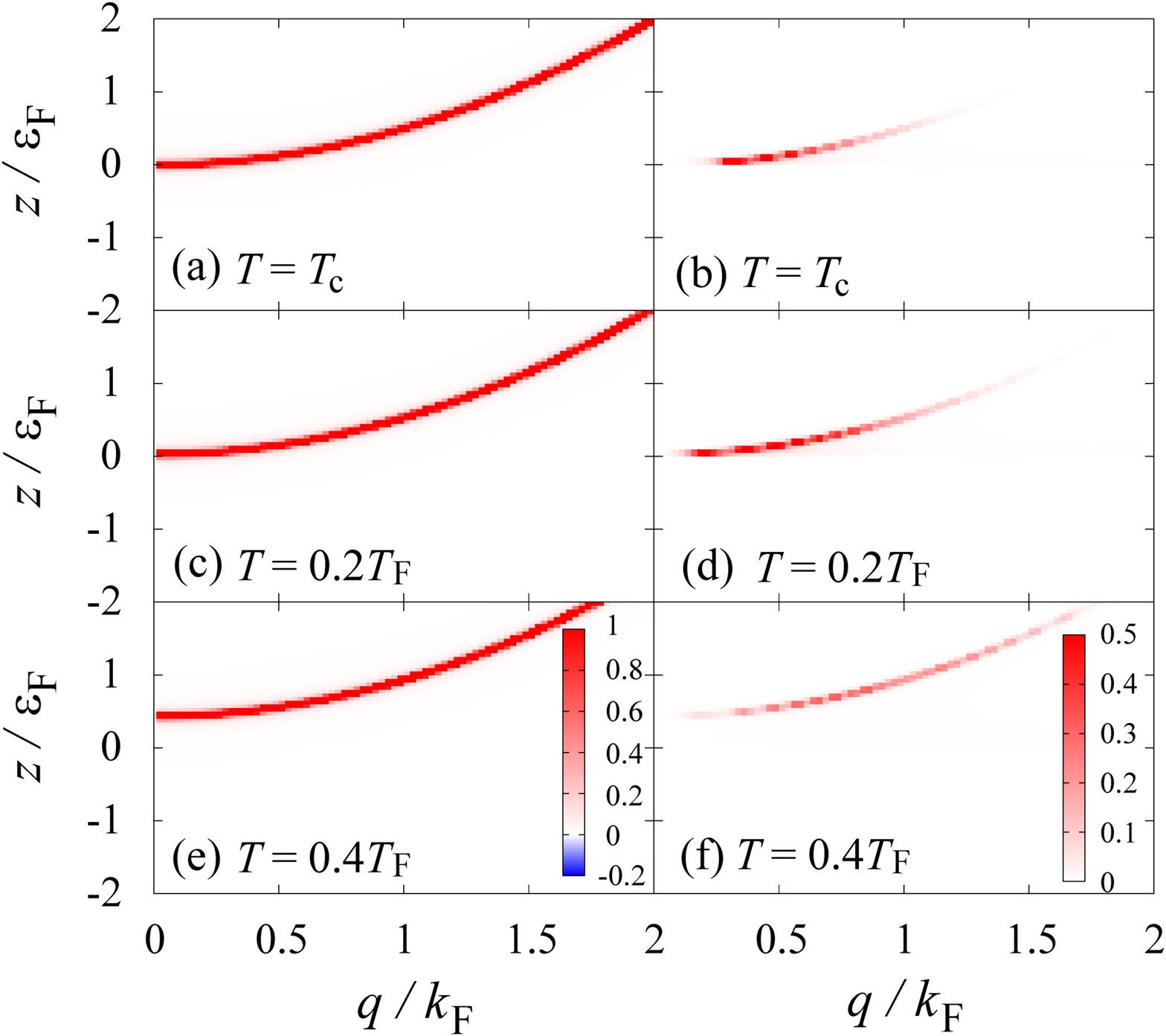}}
\caption{(Color online) Same as Fig. \ref{fig6} for the strong-coupling regime $(v_zk_{\rm F}^3)^{-1}=20$.}
\label{fig7}
\end{figure}

On the other hand, as shown in Fig. \ref{fig7}, in the strong-coupling regime, the change of the structure of ${\rm{Im}} \left[ \Gamma_{z,z} \left({\bm q}, i\nu_n \to z+i\delta \right)\right]$ is always dominant in the whole temperature regime. In other words, the thermally dissociation of the preformed Cooper pairs is the most crucial finite temperature effects in the BEC regime. Thus, $\bar{C}_v$ and $\bar{C}_R$ monotonically decreases as decreasing the number of the molecular boson with increasing $T$.
\par
\par
\par
\par
In Fig. \ref{fig5}, we also plot $C_v^{(i)}$ and $C_R^{(i)}$. We find that the weaker-coupling components ($i=x,y$) are gradually enhanced with increasing $T$. When we simply consider  the strong-coupling regime, the difference of the binding energy $\delta E_{\rm{bind}}$ between the stronger-coupling $p_z$-wave molecules and the weaker-coupling $p_x$ and $p_y$-wave molecules is given by $\delta E_{\rm{bind}}=R/(m\delta v)$. In the low temperature regime where $T \ll \delta E_{\rm{bind}}$, since most of Fermi atoms form the $p_z$-wave pairs, $C_v^{x}=C_v^{y}$ and $C_R^{x}=C_R^{y}$ are negligibly small. However, as increasing $T$, the thermal transfer from the $p_z$-wave molecule to the $p_x$ and $p_y$-wave molecule occurs. Then the difference between $C_v^{(z)}$ and $C_v^{(x)}=C_v^{(y)}$ (or $C_R^{(z)}$ and $C_R^{(x)}=C_R^{(y)}$) gradually becomes small, and eventually vanishes in the high temperature region where $T \gg \delta E_{\rm {bind}}$. Although this effects are clearly seen when the $\delta E_{\rm{bind}}$ is comparable to the Fermi energy, unfortunately, in $^{40}$K Fermi gas $\delta E_{\rm {bind}}$ is much larger than the Fermi energy. Thus it is difficult to observe this phenomenon, which essentially comes from the anisotropy of the $p$-wave interaction, in this system. 
\par 
\begin{figure}
\centerline{\includegraphics[width=8cm]{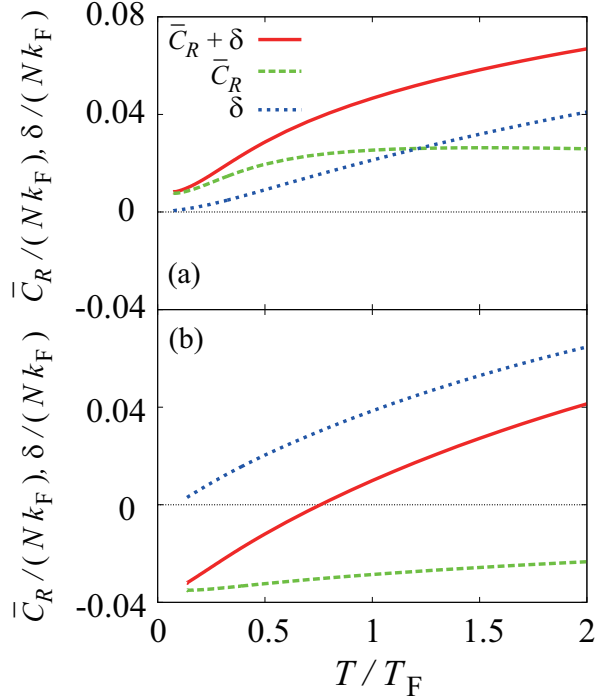}}
\caption{(Color online) Comparison between the temperature dependence of the coefficient $\bar{C}_{R}$ and $\delta$ of the sub-leading term of the asymptotic form of the momentum distribution in Eq. (\ref{eq29}). In panel (a) and (b), the results in the weak-coupling regime $(v_zk_{\rm F}^3)^{-1}=-20$ and the strong-coupling regime $(v_zk_{\rm F}^3)^{-1}=20$ are shown, respectively [$(\delta v k_{\rm F}^3)^{-1}=8$].}
\label{fig8}
\end{figure}
\par
In Fig. \ref{fig8}, we compare $\bar{C}_R$ with $\delta$ in Eq. (\ref{eq29}) as a function of $T$. As mentioned above, $\delta$ comes from the center-of-mass (c.m) motion of the pair molecules. Indeed, in high temperature region since the pairs are easily excited to the states with finite c.m momentum $q$, $\delta$ becomes remarkable, and when $T \gesim T_{\rm F}$ the coefficient of the sub-leading term of $\bar{n}(p_x,p_y)$ is eventually dominated by $\delta$ , compared to $\bar{C}_R$. This results implies that in order to measure the $p$-wave contact $\bar{C}_R$, it is needed to sufficiently decrease the temperature ($T \simeq T_{\rm c}$) or $C_R^{(i)}$ should be directly estimated from the thermodynamic properties of the system.
\section{Comparison with the experimental value of the $p$-wave contacts}
\label{trap}
Before showing the results in the trapped system, we first summarize our detailed numerical parameter setting in the calculations. Following Ref. \cite{Luciuk}, we take the total atomic number $N=36000$ and the trap frequency $\omega=440{\rm Hz}$. The scattering volume $v_i$ is estimated from the formula
\begin{equation}
v_i=v_i^{\rm{bg}} \left(1-\frac{\Delta_i}{\delta B_i} \right)
\label{eq44}
\end{equation}
where $\delta B_i=B-B_{\rm {res}}^{(i)}$ ($B_{\rm {res}}^{(i)}$ is the resonance field in the $p_i$-wave channel). For $v_i^{\rm{bg}}$, $\Delta_i$ and $B_{\rm {res}}^{(i)}$, we use the values in Ref. \cite{Ticknor}. In the following, the results are shown as a function of $\delta B_z$. For simplicity, we ignore the magnetic field dependence of the effective range. However, we mention that, in the region where the calculation is done, $R$ changes only about 2 percent.
\par
\begin{figure}
\centerline{\includegraphics[width=8cm]{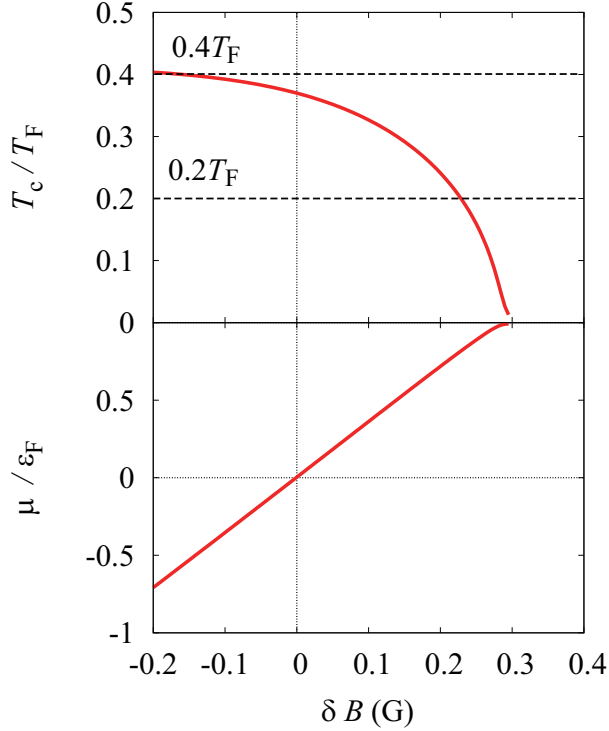}}
\caption{(Color online) phase diagram of a $p$-wave interacting Fermi gas with a harmonic trap potential and chemical potential at the trap center on $T_{\rm c}$, calculated within the local density approximation. As shown from this figure, at $T=0.2T_{\rm F}$ the superfluid transition occurs in the weak-coupling regime ($\delta B \simeq 0.275$G).}
\label{fig9}
\end{figure}
\par
Figure \ref{fig9} shows the phase diagram of a one-component Fermi gas with a $p$-wave interaction and the chemical potential at $T_{\rm c}$ calculated within the LDA. Although the validity of the LDA for $T_{\rm c}$ and $\mu$, as well as the $p$-wave contacts is still unclear in the $p$-wave case, in the $s$-wave case, the experimental results on the Tan's contact is well-described within the LDA. In Ref. \cite{Luciuk}, it was mentioned that the experiment has been done at $T=0.2T_{\rm F}$ and the system is in the normal phase. However, within the LDA, at $T=0.2T_{\rm F}$, the superfluid transition occurs at $\delta B \simeq 0.275{\rm G}$ in the weak-coupling regime.
\par
\begin{figure}
\centerline{\includegraphics[width=8cm]{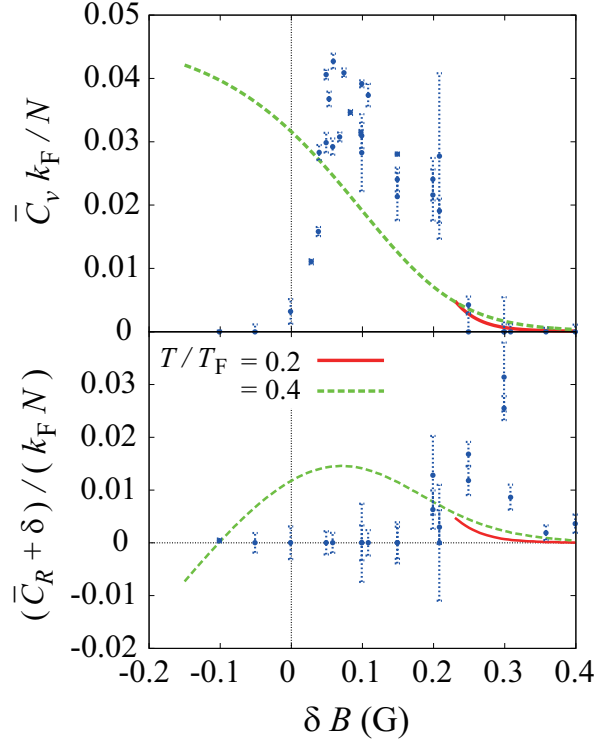}}
\caption{(Color online) calculated $\bar{C}_v$ and $\bar{C}_R+\delta$ at $T=0.2T_{\rm F}$ (solid line) and $0.4T_{\rm F}$ (dashed line) as a function of the external magnetic field. The filled square with the error bar shows experimental results from Ref. \cite{Luciuk}.}
\label{fig10}
\end{figure}
\par
\begin{figure}
\centerline{\includegraphics[width=8cm]{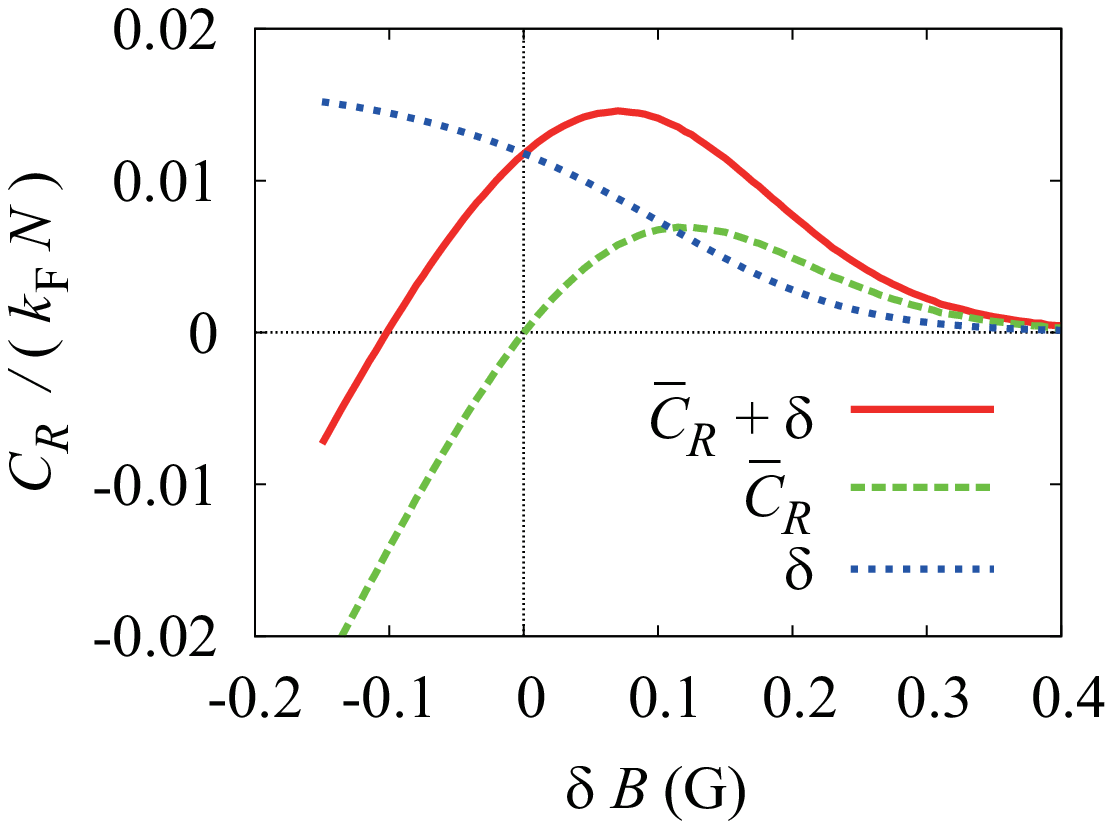}}
\caption{(Color online) Comparison between $\bar{C}_R$ and $\delta$ in $^{40}$K Fermi gas with the harmonic trap potential at $T=0.4T_{\rm F}$.}
\label{fig11}
\end{figure}
\par
In figure \ref{fig10}, we compare the calculated $\bar{C}_v$ and $\bar{C}_R+\delta$ with the experimental results. In addition to the result at the experimental temperature $T=0.2T_{\rm F}$, we also plot the results at $T=0.4T_{\rm F}$, in order to compare in the intermediate and the strong-coupling region. Experimentally, in the region where $\delta B < 0$, the $p$-wave molecules are unstable. Thus the $p$-wave contacts rapidly decrease around $\delta B=0 {\rm G}$. In our model, since the finite life time of the $p$-wave molecules does not considered, even in this regime $\bar{C}_v$ and $\bar{C}_R+\delta$ have finite value. Our calculation describes the characteristic behavior of $\bar{C}_v$ and $\bar{C}_R+\delta$. For example, the increase of $\bar{C}_v$ as a function of $\delta B$ and a peak structure appearing in $\delta B$ dependence of $\bar{C}_R+\delta$. However, our results cannot qualitatively describe the experimental results. In Fig. \ref{fig11}, we also compare $\bar{C}_R$ with $\delta$ in the trapped system. We find that even at $T=0.4T_{\rm F}$, the contribution from $\delta$ cannot be ignored in the strong-coupling regime.
\par
\begin{figure}
\centerline{\includegraphics[width=8cm]{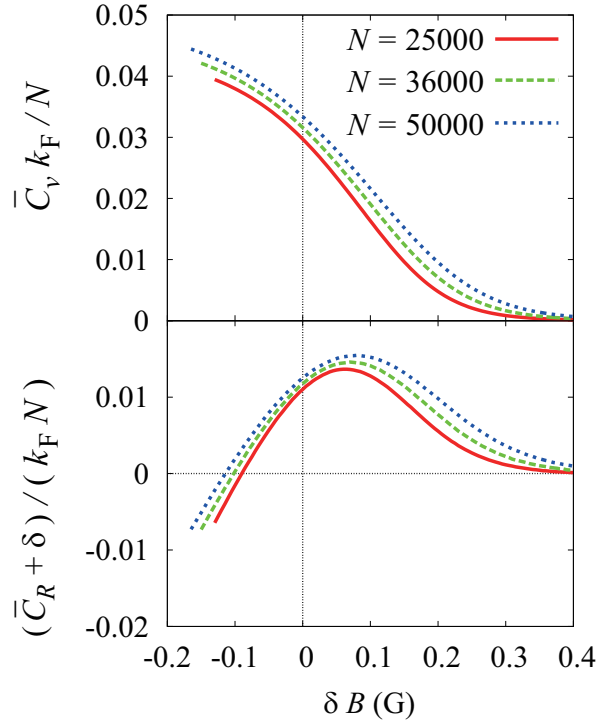}}
\caption{(Color online) particle number dependence of calculated $\bar{C}_v$ and $\bar{C}_R+\delta$ at $T=0.4T_{\rm F}$. We show the results for $N=25000$, $36000$ (experimental situation in Ref. \cite{Luciuk}) and $50000$.}
\label{fig12}
\end{figure}

Finally, we mention that the $p$-wave contacts explicitly depend on the atomic number $N$. Figure \ref{fig12} shows the $p$-wave contacts as a function of $\delta B$ for three different particle number. The $N$-dependence of $\bar{C}_v$ and $\bar{C}_R+\delta$ originates from the fact that the $p_i$-wave interaction is essentially described by two parameters, i.e. $(v_i k_{\rm F}^3)^{-1}$ and $Rk_{\rm F}$. Experimentally $v_i$ and $R$ are tuned with varying the external magnetic field and the Fermi momentum $k_{\rm F}$ is determined from the total atomic number $N$. Thus, in order to fix both $(v_i k_{\rm F}^3)^{-1}$ and $Rk_{\rm F}$, we have to tune not only the external magnetic field and but also the atomic number. As a result, even when $\bar{C}_v$ and $\bar{C}_R+\delta$ is renormalized by the atomic number, the value of these quantities as a function of the magnetic field still depends on the atomic number. Thus, to precisely measure the $p$-wave contacts, one needs to do the measurement with fixing the atomic number in addition to tuning of the external magnetic field.

\par
\par

\par
\par
\section{Summary}
To summarize, including the effects of the $p$-wave pairing fluctuations within the NSR theory, we theoretically investigated strong-coupling and finite temperature effects on the $p$-wave contacts. We found that while the $p$-wave contact related to the scattering volume simply increases as increasing the interaction strength, one related to the effective range non-monotonically depends on the interaction strength and its sign changes in the intermediate regime. We clarified that this non-monotonic behavior originate from the competition between the increase of the cutoff momentum and the decrease of the coupling constant with increasing the effective range. We also investigate the temperature dependence of the $p$-wave contacts, as well as the contribution from the extra terms which is not related to the thermodynamic properties. We showed that in high temperature regime the sub-leading behavior of the large-momentum distribution is dominated by the extra terms. This results imply that in order to measure the $p$-wave contacts, one need to decrease the temperature near the superfluid transition temperature $T_{\rm c}$ or directly observe the thermodynamic quantities of the system.
\par
By including the effects of the trap potential within the local density approximation, we directly compared our results with the recent experiment. Although in quantitative level, the experimental results can not be described by our model, we obtained similar magnetic field dependence of the $p$-wave contacts. We also mentioned that the $p$-wave contacts explicitly depends on the atomic number due to the fact that the $p$-wave interaction is characterized by two parameters. Thus we concluded that in order to precisely measure the $p$-wave contacts one needs not only to tune the external magnetic field but also to control the particle number of the system. In the current stage of cold atom physics, it is one of the most challenging issues to understand how the $p$-wave pairing fluctuations affects the many-body properties of the system. Since the $p$-wave contacts connect the microscopic properties with the thermodynamic properties, our results are useful for further understanding the physical properties in the $p$-wave interacting Fermi gas in the normal phase.
\par
\par
\begin{acknowledgments}
This work was supported by KiPAS project in Keio University. DI was supported by Grant-in-aid for Scientific Research from JSPS in Japan (No.JP16K17773). YO was supported by Grant-in-aid for Scientific Research from MEXT and JSPS in Japan (No.JP15H00840, No.JP15K00178, No.JP16K05503). 
\end{acknowledgments}
\par

\end{document}